\documentstyle[aps,multicol,epsf]{revtex}
\begin{document}
\draft
\title{Diffusion and viscosity in a supercooled polydisperse system} 

\vspace{1cm}
\author{Rajesh K. Murarka and Biman Bagchi\footnote[1]{For correspondence: bbagchi@sscu.iisc.ernet.in}}
\address{Solid State and Structural Chemistry Unit, \\
Indian Institute of Science, \\
Bangalore, India 560 012.}
\maketitle

\begin{abstract}

We have carried out extensive molecular dynamics simulations of a supercooled polydisperse
Lennard-Jones liquid with large variations in temperature at a fixed pressure. The particles 
in the system are considered to be polydisperse both in size and mass. The temperature 
dependence of the dynamical properties such as the viscosity ($\eta$) and the self-diffusion 
coefficients ($D_i$) of different size particles is studied. Both viscosity and diffusion 
coefficients show super-Arrhenius temperature dependence and fit well to the 
well-known Vogel-Fulcher-Tammann (VFT) equation. Within the temperature range investigated, the
value of the Angell's fragility parameter (D $\approx 1.4$) classifies the present system into a 
strongly fragile liquid. The critical temperature for diffusion ($T_o^{D_i}$) increases with the 
size of the particles. The critical temperature for viscosity ($T_o^{\eta}$) is larger than that 
for the diffusion and a sizeable deviations appear for the smaller size particles implying a decoupling of 
translational diffusion from viscosity in deeply supercooled liquid. Indeed, the diffusion shows markedly 
non-Stokesian behavior at low temperatures where a highly nonlinear dependence on size is observed. 
An inspection of the trajectories of the particles shows that at low temperatures the motions of 
both the smallest and largest size particles are discontinuous (jump-type). However, the crossover from 
continuous Brownian to large length hopping motion takes place at shorter time scales for the smaller size 
particles.
      
\end{abstract}
  
\begin{multicols}{2}
\section{Introduction}

The rapid cooling of a liquid below its freezing temperature transforms it into 
a long-lived metastable amorphous solid or glass\cite{angel1}. Understanding the dynamics 
of the system near the glass transition is an intense field of research since
the last few decades. There have been many experimental\cite{ediger,weitz,kegel,vandenbout,fayer,angel2} as 
well as simulation studies\cite{kob1,sastry,glotzer,mountain,heuer,schober,perera1} which focus on the 
dynamics of dense supercooled liquids well above the glass transition and also near the glass transition 
temperature ($T_g$). The basic aim of all these studies was to characterise quantitatively the 
observed very complex dynamics of the system as it approaches the glass transition 
from the above. Close to the glass transition, the shear viscosity ($\eta$) and the microscopic 
structural relaxation time ($\tau$) of the so-called fragile glass-forming liquids show 
divergence with a strongly non-Arrhenius temperature dependence\cite{angel1}. This divergence is often 
well represented by the Vogel-Fulcher-Tammann (VFT) equation
\begin{equation}
\eta(T)\;=\;A_{\eta}\,\exp[E_{\eta}/(T - T_o^{\eta})]
\end{equation}
\noindent
where $A_{\eta}$ and $E_{\eta}$ are temperature-independent constants and $T_o^{\eta}$ ($< T_g$) is the
temperature at which $\eta$ diverges. Note that at low temperatures the increasingly 
slow dynamics of the so-called fragile liquids is simultaneously manifested in the stretched exponential 
decay of the stress correlation function (with a strongly temperature dependent stretching 
parameter)\cite{angelani1,mukherjee}. 
The VFT dependence (Eq. 1) is thus accompanied by the strong non-exponential relaxation observed near 
the glass transition temperature.    

The dramatic slow down of the dynamics near the glass transition is not well understood 
and still remains the most challenging problem in the physics of glasses. Several theories
have been proposed to understand the anomalous relaxation dynamics of deeply supercooled liquids. 
Although the ideal version of the non-linear mode coupling theory (MCT)\cite{gotze} gives a microscopic 
picture of this slowing down, it predicts a structural arrest, i.e., a transition from ergodic to 
nonergodic behavior, at a critical temperature $T_c$, well above the laboratory glass transition 
temperature $T_g$. Near $T_c$, the importance of the influence of potential energy landscape 
on the relaxation processes has now widely been accepted\cite{schroder,angelani2} and the strongly correlated 
jump motion is observed to be the dominant mode for mass transport\cite{schober,hansen,sarika1,sarika2}, 
which is not included in ideal MCT.     

Another important characteristic feature in the dynamics of deeply supercooled liquids is the 
decoupling between translation diffusion and the shear viscosity of the 
medium\cite{stillinger,kivelson,liu}. At high temperature, over a wide range of liquid states, the 
translational diffusion is inversely proportional to viscosity, in accordance to the 
Stokes-Einstein (SE) relation given by
\begin{equation}
D_T\;=\;{\frac{k_BT}{C\pi\eta R}}
\end{equation}
\noindent
where $R$ is the spherical radius of the diffusing particle and $C$ is a numerical constant that 
depends on the hydrodynamic boundary condition. However, several recent 
experimental\cite{rossler,cicerone,sillescu} and simulation 
studies\cite{angelani1,thirumalai,yamamoto1,yamamoto2,perera2,leporini} on strongly 
supercooled 'fragile' glass forming liquids have shown significant deviations from the SE relation. 
As the temperature is lowered toward $T_g$, it is found that the translational diffusion is larger than 
the value predicted by the SE relation and in some cases even two to three orders of magnitude 
larger\cite{stillinger}. 
The enhanced diffusion at low temperatures is sometimes explained in terms of a power law behavior 
$D_T \propto \eta^{-\alpha}$ with $\alpha < 1$\cite{sillescu,yamamoto1}. 
Both the experiments and simulation studies have recently been evidenced the enhancement of the 
translational diffusion coefficient is due to the spatially heterogeneous dynamics in deeply 
supercooled liquids\cite{ediger,cicerone,sillescu,yamamoto2,perera2}.   

Computer simulations have played a key role in augmenting our understanding of various aspects of 
the dynamics of the supercooled liquids from a microscopic viewpoint. Unfortunately, the simple
one-component systems such as soft or hard spheres or Lennard-Jones systems crystallize rapidly 
on lowering the temperature below the melting point ($T_m$) and, therefore, cannot be utilized
as a model for studying the complex dynamical behavior near the glass transition temperature.
A natural way to avoid crystallization is to use binary mixtures of atoms with different 
diameters. A large number of molecular dynamics (MD) simulations have recently been carried out in
supercooled model binary mixtures near the glass transition as well as below the glass transition
temperature\cite{kob1,sastry,glotzer,mountain,heuer,mukherjee,sarika1,sarika2,barrat,wahnstrom,kob2}.

However, one is often interested in the consequences of the disorder introduced by the dissimilarity 
of the particles. Synthetic colloids, by their very nature, frequently exhibit considerable size 
polydispersity\cite{pusey,russel}. Polydispersity is also common in industrially produced polymers 
which always contain macromolecules with a rang of chain length. Colloidal particles are an excellent 
model of hard spheres and perhaps the simplest possible experimental system of interacting particles to 
study the glass transition. Several experiments\cite{pusey} and simulations\cite{kofke,lacks} have 
shown that the crystal phase of the colloidal systems can exist as a thermodynamically stable phase only 
for polydispersities (standard deviation of the size distribution divided by the mean) less than 
a 'terminal' value, in the range of 0.05-0.15.
 
Interestingly, recent experiments on colloidal supercooled fluids and colloidal
glasses allowed one to obtain the information on the microscopic details of the dynamics of the 
individual particles\cite{weitz}. These experiments have shown the presence of dynamic heterogeneity in 
deeply supercooled colloidal systems. The motion of the relatively fast moving particles is found to 
be highly correlated and form connected clusters, whose size increases as one approach the glass 
transition. More recently, Sear\cite{sear} has carried out MD simulation of a dense polydisperse hard sphere
fluid to study the effect of polydispersity on the slow dynamics. The simulation results also show
the clustering of the fast-moving particles in agreement with the experiments, though the dynamics 
appears to be less heterogeneous. The heterogeneous nature of the dynamics has also been observed 
in the Monte Carlo simulation study of polydisperse hard spheres close to the glass 
transition\cite{doliwa}. 

It is worth mentioning that besides the short-range hard-core interaction, addition of a nonadsorbing 
soluble polymer in the stable colloidal suspension gives rise to a weak, long-range attraction between 
the colloidal particles by means of the depletion interaction\cite{russel}. Recently, the consequences of 
this attractive interaction on the glass transition are nicely explained in a combined experimental, 
theoretical, and simulation study by Pham {\it et.al.}\cite{pham} Interestingly, with increase in the 
strength of the short-range attractive interaction, two qualitatively different glassy states are found 
with a reentrant glass transition line. 

The size distribution in real colloids generally leads to a distribution in mass of the particles. The 
importance of the mass polydispersity on the dynamics of a realistic system having size polydispersity 
has recently been analyzed in molecular dynamics simulation study of a Lennard-Jones (LJ) 
polydisperse fluid near the triple point of the corresponding monodisperse LJ system\cite{kiriushcheva}.  
Polydispersity is commonly found in many systems of industrial applications and to mimic the 
interparticle interactions, the Lennard-Jones potential generally serve as a good starting model. 
Thus, it will be a general interest to study the impact of polydispersity on the dynamics of a 
deeply supercooled polydisperse fluid, where particles interacting via the LJ potential. 
More importantly, this will enable us to compare the properties of the system with model binary 
LJ mixtures whose dynamics near the glass transition has been studied extensively in simulations.
 
In this study, we have performed extensive molecular dynamics simulations of a system of polydisperse 
LJ spheres with continuous range of diameters and mass. The temperature dependence of the dynamic 
properties such as viscosity ($\eta$) and the self-diffusion coefficients ($D_i$) for the different 
size particles is studied by varying the temperature ($T$) over a large range at a constant high 
pressure ($P$). Both the viscosity and diffusion show super-Arrhenius temperature dependence and the 
calculated value of the fragility parameter (D) show that the present system is more fragile than 
the well-known Kob-Andersen binary mixture. The critical temperature obtained from the VFT fit to the 
diffusions ($T_o^{D_i}$) show strong dependency on the radius ($R$) of the particles. In addition, the 
critical temperature obtained from the VFT fit to the viscosity ($T_o^{\eta}$) is much higher when 
compared to those of diffusion coefficients, where the deviation is largest for the smaller size particles. This clearly reflects the deviation from the Stokesian diffusion in the proximity of the glass 
transition temperature. Most interestingly, at lower temperatures the diffusion shows
a highly nonlinear size dependence when plotted against the inverse of the radius ($R$) of the 
particles. The reason for the breakdown of the Stokes-Einstein (SE) relation can be analyzed from the 
trajectories of the particles. We find that at low temperature the hopping 
processes being the primary mode of particle diffusion for both the smaller and bigger size particles.  

The organization of the rest of the paper is as follows. In the next section, we describe in detail the 
system studied here and the details of the simulations. The simulation results are analyzed and discussed in section III. Finally, we end with some concluding remarks in section IV.
          
\section{System and Simulation Details}

We have performed a series of equilibrium isothermal-isobaric 
ensemble (N P T) molecular dynamics simulations in three dimensions of a 
system of $N = 256$ particles of mean radius ${\bar \sigma}$ and mass ${\bar m}$ with 
polydispersity in both size and mass. 
The interaction between any two particles is modeled by means of shifted force Lennard-Jones (LJ) pair potential, where the standard LJ is given by\cite{allen}
\begin {equation}
u_{ij}^{LJ} = 4 \epsilon _{ij} \left [{\left(\sigma_{ij} \over r_{ij}\right )}^{12} - {\left (\sigma_{ij} \over r_{ij} \right )}^6 \right ]
\end {equation}
\noindent where $i$ and $j$ denotes two different particles and $\sigma_{ij} = (\sigma_{i} + 
\sigma_{j})/2$ where $\sigma_{i}$, $\sigma_{j}$ are the diameters of the particles 
$i$ and $j$, respectively. In the shifted-force potential both the potential and force are
continuous at a cutoff radius $r_c$ and we choose a value of $r_c = 2.5{\bar \sigma}$.
 
The polydispersity in size is introduced by random sampling from the 
Gaussian distribution of particle diameters $\sigma$\cite{kiriushcheva}
\begin{equation}
P(\sigma)\;=\;{\frac{1}{\delta\sqrt{2\pi}}}\exp\Biggl[-{1\over 2}\Biggl({{\sigma - {\bar \sigma}}\over \delta}\Biggr)^2\Biggr],
\end{equation}
\noindent where $\delta$ is the width of the distribution. The standard deviation ($\delta$) of the 
distribution divided by its mean ${\bar \sigma}$ gives a dimensionless parameter, the polydispersity 
index $s = \delta/{\bar \sigma}$. The simulations are carried out here for a fixed value of the 
polydispersity index $s = 0.1$. The masses of the particles are varied accordingly and we assume that 
the mass of a particle $i$ is scaled by its diameter as $m_i = {\bar m}(\sigma_{i}/{\bar \sigma})^3$.
The LJ energy parameter $\epsilon_{ij}$ is assumed to be same for all particle pairs and denoted as  
$\epsilon$. All the quantities in this study are given in reduced units, that is, length in units 
of ${\bar \sigma}$, temperature $T$ in units of $\epsilon/k_B$, pressure $P$ in units of 
$\epsilon/{\bar \sigma}^{3}$, and the unit of time is $\tau = \sqrt{{\bar m}{\bar \sigma}^2/\epsilon}$. 
Note that if one assumes the argon units then $\tau$ = 2.2 ps.

All simulations in the NPT ensemble were performed using the Nose-Hoover-Andersen method\cite{nose}, where 
the external reduced temperature ($T^*$) is varied over a large range from 1.3 to 0.67 keeping the 
external reduced pressure ($P^*$) fixed at 10.0. Throughout the course of the simulations, the 
barostat and system's degrees of freedom are coupled to an independent 
Nose-Hoover chain\cite{martyna1} (NHC) of thermostats, each of length 5. The extended system 
equations of motion are integrated using the reversible integrator method\cite{tuckerman}.
The higher order multiple time step method has been 
employed in the NHC evolution operator which lead to stable energy conservation for 
non-Hamiltonian dynamical systems\cite{martyna2}. The extended system time scale parameter 
used in the calculations was taken to be $0.93$ for $T^* \ge 1.0$ and $1.16$ for 
$T^* < 1.0$ for both the barostat and thermostats.   
   
A time step of $0.001$ is employed for $T^* \ge 1.0$ and $0.002$ for $T^* < 1.0$. 
The equilibration and data collection steps are also varied accordingly depending upon the temperature 
of the system. For $T^* \ge 1.0$, the equilibration steps are varied from $5 \times 10^5$ to 
$10^6$ and the data collection steps are $10^6$, whereas for $T^* < 1.0$, the equilibration 
steps are varied from $5.0 \times 10^5$ to $2.0 \times 10^6$ and the data collection steps 
from $10^6$ to $2.5 \times 10^7$. At each temperature, all the dynamic quantities are 
averaged over {\it five} independent runs. Diffusion coefficients ($D_i$) for the different size 
particles are calculated from the slope of the corresponding mean square displacements (MSD) in the 
diffusive limit and viscosity is calculated from the auto-correlation of 
the off-diagonal components of the microscopic stress tensor, via the standard Green-Kubo 
formula\cite{hm}. 
As the system is isotropic, we have taken an average over three different off-diagonal stress 
correlations for each of the five data set.    

\section{Results and Discussion} 

In order to make sure that there is no crystallization, we have calculated the radial distribution 
functions $g(r)$ which describes the average structure of the fluid. The radial distribution function 
calculated for $T^* = 0.67$ is shown in figure 1. The decay of correlations with increase in distance 
clearly reflects the absence of any long range order, a characteristic feature of the fluid.

The plot of ln($\eta$) as a function of inverse of temperature ($1/T^*$) in figure 2(a) clearly 
shows a super-Arrhenius behavior of the viscosity. In figure 2(b) we show a VFT fit to the 
viscosity (Eq. 1) by plotting ln($\eta$) against (1/($T - T_o^{\eta}$)) where 
$T_o^{\eta}$ is equal to 0.57. As in other fragile liquids, it shows that the divergence of viscosity 
is quite well described by the VFT equation. From the fitting we obtain the values of $A_{\eta}$ and 
$E_{\eta}$ as 2.0 and 0.81, respectively. We have also calculated the fragility parameter (D 
= $E_{\eta}/T_o^{\eta}$) as defined by Angell\cite{angel3}. Using the values of the fitting parameters 
($E_{\eta}$ and $T_o^{\eta}$) obtained within the temperature range investigated, its value is 
$\approx 1.42$. This classifies the present system into a strongly fragile liquid. Thus, the dense 
random packing of unequal size particles makes the present system more fragile when compared with a recent 
simulation study on Kob-Andersen binary mixture (D $\approx 2.45$)\cite{mukherjee}.
              
In a polydisperse system, all the particles are unequal in size, so their diffusion coefficients 
will also differ. We categorize the particles into different subsets where particles of 
diameters within $0.05{\bar \sigma}$ are assumed to be members of the same subset. For the 
polydispersity index $s = 0.1$, we find that the minimum and maximum diameter of the particles 
are $0.75{\bar \sigma}$ and $1.25{\bar \sigma}$, respectively. Thus subsets of particles with 
diameters in the ranges $0.75$ to $0.8{\bar \sigma}$ and $1.2$ to $1.25{\bar \sigma}$ corresponds 
to smallest and largest spheres, respectively. The diffusion coefficients for different subsets 
of particles are calculated at each temperature. 
It is well-known that in deeply supercooled liquids the diffusion coefficient shows non-Arrhenius 
temperature dependence and can be fitted by a VFT law 
\begin{equation}
D_i(T)\;=\;A_{D_i}\,\exp[-E_{D_i}/(T - T_o^{D_i})],
\end{equation}
\noindent where the index $i$ stands for the different subsets of particles. $T_o^{D_i}$ is the 
critical temperatures for $i$-th species at which the diffusion coefficients ($D_i$) vanishes. 
The diffusion coefficients for each subsets of particles have been fitted to the above equation 
and we show the plot of ln($D_i$) against $1/(T - T_o^{D_i})$, for the smallest ($D_1$) and largest 
spheres ($D_{10}$) in figures 3(a) and 3(b), respectively, where $T_o^{D_1}$ and $T_o^{D_{10}}$ are 
0.46 and 0.5. Thus the critical temperatures ($T_o^{D_i}$) depends on the size of the particles and 
increases with size of the particles ${{\bar R}_i}$ (${{\bar R}_i}$ is the mean radius of the 
$i$-th subset). This is shown in figure 4. For the largest spheres, the critical 
temperature ($T_o^{D_{10}}$) is smaller than the corresponding critical temperature obtained from 
VFT fit to the viscosity ($T_o^{\eta} = 0.57$). This clearly signifies that near the glass transition 
the diffusion is partly decoupled from the viscosity and for smaller particles the degree of decoupling 
is more. The smaller particles remain mobile even when bigger particles are almost frozen. 

In figure 5 we plot the diffusion constants ($D_i$) against $1/{{\bar R}_i}$ at the lowest temperature 
of $T^* = 0.67$ and compared with the well-known hydrodynamic Stokes-Einstein (SE) relation (Eq. 2 with 
$C$ = 6, the stick boundary condition). It clearly shows the markedly non-Stokesian behavior of the 
diffusion at low temperatures. Interestingly, the fitting to the simulated data points show a highly 
nonlinear size dependence of the diffusion. This is a clear evidence that the breakdown of SE law is 
more severe for the smaller size particles. In order to get an estimate of the degree of decoupling 
(between diffusion and viscosity) for the smallest size particles, we have fit the inverse diffusion 
coefficient ($1/D_1$) versus ${\eta}/T$. While at high T, it asymptotically satisfies the SE relation 
(slope is 1), the fit to the low temperature data gives the slope $\alpha \approx 0.5$ (that is, diffusion 
shows the power law behavior $D_1 \propto \eta^{-0.5}$). It should be noted that the dynamics of a 
polydisperse liquid is more heterogeneous than a monodisperse or bidisperse system due to the different 
time scales involved for different sizes and masses of the particles\cite{sear,kiriushcheva}. The smaller 
particles are on average faster than others over all time scales. This becomes more prominent at lower 
temperatures where the relaxation time of the system is very high. At low temperature, the observed 
nonlinear dependence of diffusion on size is related to the increase in dynamic heterogeneity in a 
polydisperse system.

A more detailed analysis of the diffusion can be obtained from a closer examination of the 
self-part of the van Hove correlation function $G_s(r,t)$. This gives the distribution of the 
displacements ($r$) of a particle in a time interval $t$. We calculate $G_s(r,t)$ for the smallest 
($\sigma$ = 0.75 to 0.8${\bar \sigma}$) and largest ($\sigma$ = 1.2 to 1.25${\bar \sigma}$) particles 
for different time intervals at $T^* = 0.67$, the lowest temperature investigated. Figures 6(a) and 
6(b) shows the correlations for the smallest and largest spheres, respectively. For the smallest particles, a gradual development of a well-defined second peak at $r \sim 1.0{\bar \sigma}$ is clearly visible with 
increase in time (figure 6(a)). However, for the largest particles, the distribution becomes bimodal at 
relatively longer time scales (figure 6(b)). 
The occurrence of the secondary peak, observed also in other model binary mixtures at low 
temperatures\cite{schroder,barrat,wahnstrom}, is an evidence of the jump motion in the dynamics of the 
particles.

To characterize the single particle dynamics further, we have evaluated the self-intermediate 
scattering function $F_s(k,t)$, the spatial Fourier transform of $G_s(r,t)$, for different subsets of 
the particles for a fixed value of the reduced wave number $k{\bar \sigma} \sim 2\pi$ at $T^* = 0.67$. 
The long time decay of $F_s(k,t)$ is well fitted by the Kohlrausch-Williams-Watts (KWW) stretched 
exponential form
\begin{equation}
F_s^i(k,t)\;=\; \exp\Biggl(-{t \over \tau_i}\Biggr)^{\beta_i}
\end{equation}
\noindent where $\tau_i$ and $\beta_i$ are the relaxation time and the stretching exponent of the 
$i$-th subset. We find that both $\tau_i$ and $\beta_i$ increases 
with an increase in the size of the particles, as has been observed earlier by other authors in binary
mixtures\cite{perera1}. The $F_s(k,t)$ calculated for the smallest (subset 1) and 
largest (subset 10) particles along with the KWW fits is shown in figures 7(a) and 7(b), respectively. 
Note that we fit the functions $[F_s^i(k,t-t_o)/F_s^i(k,t_o)]$ ($t > t_o$), to the KWW form to 
quantify their long time behavior\cite{sastry}. For the smallest particles, the values of the fitting 
parameters are found to be $\tau_1 \simeq 242$ and $\beta_1 \simeq 0.49$, whereas for the largest particles 
they are $\tau_{10} \simeq 717$ and $\beta_{10} \simeq 0.64$. The enhanced stretching 
($\beta_1 < \beta_{10}$) at long times is due to the more heterogeneity probed by the smaller size 
particles than that by the larger size particles during the time scale of decay of 
their $F_s(k,t)$\cite{murarka}.
              
In order to determine the extent of the jump-type motion more clearly we follow the trajectory of the 
individual particles. A close inspection of the simulated trajectory of the smallest and largest particles 
reveals several interesting features. Figure 8(a) display the projections onto an x-y plane of the 
trajectory of a typical smallest size particle over a time interval $\Delta t = 500\tau$ and 
figure 8(b) shows the paths followed by a largest size particle over time interval $\Delta t = 2000\tau$, 
both at $T^* = 0.67$. At this temperature the dynamics is dominated by 'hopping'; particles remain 
trapped in transient cages created by the surrounding particles for quite some time and then moves 
significant distance (approximately one interparticle distance) by making a jump to a new cage. 
For the larger size particles, the jump motion begins to take place at relatively longer time 
scales (compare also figures 6(a) and 6(b)). Thus, it clearly elucidates that in a system with 
particles of all different sizes and masses, the hopping is the dominant diffusive mode both for the 
smaller and bigger size particles. It is the frequent hopping in case of smaller size particles 
that leads to the severe breakdown of SE relation.     
                
\section{Conclusions}

In summary, we have presented the results of large scale computer simulations for a supercooled 
polydisperse system with large variations in temperature at a fixed high pressure. 
The interparticle interaction is represented by the standard Lennard-Jones (LJ) potential and the 
particles in the system are considered to be polydisperse both in size and mass. Characteristic of 
a fragile glass former, the super-Arrhenius temperature dependence is observed for the viscosity 
and also for the self-diffusion coefficients of different size particles. Furthermore, within the 
temperature range investigated, the value obtained for the Angell's fragility parameter (D $\approx 1.4$) 
establish the present system as a {\it strongly fragile} liquid. 

In a dense polydisperse system, a wide range of time scales are involved due to the different size and 
mass of the particles. Thus, upon lowering the temperature, the dynamics is increasingly more 
heterogeneous compared to a monodisperse or bidisperse system. The increase of critical temperature 
for diffusion ($T_o^{D_i}$) with the size of the particles suggests that the dynamics is indeed 
heterogeneous. In addition, the critical temperature for viscosity ($T_o^{\eta}$) is found to be 
larger than that for the diffusion, indicating the decoupling of translational diffusion 
from the viscosity commonly observed in deeply supercooled liquids. Interestingly, a marked 
deviation from the Stokesian diffusion is observed where the dependence on size of the particles is 
highly nonlinear. 

The analysis of the self-part of the van Hove correlation functions $G_s(r,t)$ showed a clear 
signature of the single particle 'hopping' (indicated by the second neighbor peak) in the 
dynamics of both the smallest and largest size particles at low temperatures. However, for the 
larger particles, the hopping processes are found to occur at relatively longer times. 
The relevance of the hopping processes at low temperatures is further investigated in detail by 
following the trajectory of the individual particles. Crossover from continuous Brownian to hopping 
motion takes place at shorter time scales for the smaller size particles than that for the larger size 
particles. 

In the present system the size of all the particles are different. It would be interesting to see 
whether the jump motion executed by the individual particles occurs over a single energy barrier or 
it takes place via a number of "intermediate" inherent structures in the potential energy landscape. 
A recent molecular dynamics simulations on LJ binary mixture\cite{schroder} have shown that such a 
transition does not correspond to transitions of the system over single energy barriers. 
In addition, in a deeply supercooled liquid the jump motions are associated with strong nearest-neighbour 
correlations, in which several neighboring atoms jump at successive close 
times\cite{hansen,sarika1,wahnstrom}. It is to be noted that similar correlations have been observed here 
also. Recently, a computer simulation study of a deeply supercooled binary mixture\cite{sarika2} has shown 
that the local anisotropy in the stress is responsible (at least partly) for the particle hopping. 
However, the molecular origin of the jump motions observed here (highly disordered system) is not clear 
and we are presently pursuing this problem. 
                 
\vspace{1cm}
{\bf Acknowledgments}

This work was supported in part by the Council of Scientific and Industrial
Research (CSIR), India and the Department of atomic energy (DAE), India.
One of the authors (R.K.M) thanks the University Grants Commission (UGC) for
providing the Research Scholarship.

\newpage
 
{\large \bf Figure Captions}

{\bf Figure 1.} The radial distribution function $g(r)$ of the system at $T^* = 0.67$, the lowest temperature 
investigated. 

{\bf Figure 2.} Temperature dependence of the shear viscosity ($\eta$). (a) ln($\eta$) as a function of 
inverse of temperature ($1/T^*$). The simulating values given by the solid circles show super-Arrhenius 
behavior. The dashed line gives a guideline to the Arrhenius behavior. (b) ln($\eta$) against 
1/($T-T_o^{\eta}$). The solid circles are again represents the simulation results and the VFT fit is 
shown by the solid line. 
$T_o^{\eta}$ is found to be 0.57. The slope ($E_{\eta}$) and intercept (ln $A_{\eta}$) obtained from the fit 
are 0.81 and 0.69, respectively. For details, see the text.

{\bf Figure 3.} Temperature dependence of the diffusion coefficients for the smallest ($D_1$) and largest ($D_{10}$) particles. (a) ln($D_1$) is plotted against 1/($T-T_o^{D_1}$). (b) ln($D_{10}$) against 
1/($T-T_o^{D_{10}}$). Solid circles are the simulation results and the solid lines are the VFT fit. 
The critical temperatures $T_o^{D_1}$ and $T_o^{D_{10}}$ obtained from the VFT fit are 0.46 and 0.50, respectively. 

{\bf Figure 4.} The critical temperature ($T_o^{D_i}$) obtained from the VFT fit to the different subsets of 
particles as a function of the mean radius (${\bar R}_i$) of the subsets (in units of ${\bar \sigma}$). Note that $T_o^{D_i}$ increases with the size of the particles. 

{\bf Figure 5.} The diffusion coefficients ($D_i$) as a function of 1/${\bar R}_i$ at $T^* = 0.67$. The dashed line represents the Stokes-Einstein relation (Eq. 2) with the stick boundary condition $C = 6$. The viscosity 
($\eta$) value is taken from the present simulations. Solid circles are the simulated values and the 
solid line is the cubic polynomial fit in 1/${\bar R}_i$. The fit parameters are as follows: $D_i = 0.0011 - 0.00132(1/{\bar R}_i) + 0.000442(1/{\bar R}_i)^2$. It clearly shows a highly nonlinear size dependence and 
a marked deviation from the Stokesian behavior of the diffusion.

{\bf Figure 6.} The van Hove self-correlation function $G_s(r,t)$ as a function of the particle displacements 
$r$ (in units of ${\bar \sigma}$) at $T^* = 0.67$ for different values of time t (in units of $\tau = \sqrt{{{\bar m}{\bar \sigma}^2}/\epsilon}$ = 2.2 ps for argon units). (a) For the smallest size particles (subset 1). 
The occurrence of the second peak at $r \approx 1.0{\bar \sigma}$ indicates the single particle hopping. (b) 
For the largest size particles (subset 10). Here also a second peak corresponds to single particle hopping 
develops but at relatively longer times. 

{\bf Figure 7.} The self-intermediate scattering function $F_s(k,t)$ for $T^* = 0.67$ is shown with a shift 
in the time origin to $t_o = 1.0$, and normalized to the value at $t_o$, for a fixed value of $k{\bar \sigma} = 2\pi$. This transformation is a convenient way to eliminate the Gaussian dependence at short time\cite{sastry}. (a) For the smallest size particles (subset 1). (b) For the largest size particles (subset 10). Open circles 
represents the simulation results and the solid lines are the stretched exponential fit (Eq. 6). The time 
constants ($\tau_1$ and $\tau_{10}$) and the exponents ($\beta_1$ and $\beta_{10}$) obtained from the 
fits are $\tau_1 \simeq 242$, $\beta_1 \simeq 0.49$, $\tau_{10} \simeq 717$, and $\beta_{10} \simeq 0.64$.

{\bf Figure 8.} (a) Projections into x-y plane of the trajectory of a typical smallest size particle over a 
time interval $t = 500\tau$. (b) Projections into x-y plane of the trajectory of a typical largest size 
particle over a time interval $t = 2000\tau$. Note that the time (t) is scaled by $\tau = \sqrt{{{\bar m}{\bar \sigma}^2}/\epsilon}$ it is 2.2 ps if argon units are assumed. For detailed discussion, see the text.           
                     
\end{multicols}

\begin{thebibliography} {abcd-uf}

\bibitem{angel1} M. D. Ediger, C. A. Angell, and S. R. Nagel, J. Phys. Chem. {\bf 100}, 13200 (1996); 
C. A. Angell, K. L. Ngai, G. B. McKenna, P. F. McMillan, and S. W. Martin, J. Appl. Phys. {\bf 88}, 3113 (2000); P. G. Debenedetti and F. H. Stillinger, Nature {\bf 410}, 259 (2001). 

\bibitem{ediger} M. D. Ediger, Annu. Rev. Phys. Chem. {\bf 51}, 99 (2000).

\bibitem{weitz} E. R. Weeks, J. C. Crocker, A. C. Levitt, A. Schofield, and D. A. Weitz, Science {\bf 287}, 627 (2000).

\bibitem{kegel} W. K. Kegel and A. van Blaaderen, Science {\bf 287}, 290 (2000).

\bibitem{vandenbout} L. A. Deschenes and D. A. Vanden Bout, Science {\bf 292}, 255 (2001).

\bibitem{fayer} G. Hinze, D. D. Brace, S. D. Gottke, and M. D. Fayer, J. Chem. Phys. {\bf 113}, 3723 (2000); Phys. Rev. Lett. {\bf 84}, 2437 (2000).

\bibitem{angel2} L.-M. Martinez and C. A. Angell, Nature {\bf 410}, 663 (2001).

\bibitem{kob1} W. Kob and H. C. Andersen, Phys. Rev. Lett. {\bf 73}, 1376 (1994); Phys. Rev. E {\bf 51}, 4626 (1995); W. Kob, J. Phys.: Condens. Matter {\bf 11}, R85 (1999).

\bibitem{sastry} S. Sastry, P. G. Debenedetti, and F. H. Stillinger, Nature {\bf 393}, 554 (1998); S. Sastry, Phys. Rev. Lett. {\bf 85}, 590 (2000); Nature {\bf 409}, 164 (2001).

\bibitem{glotzer} W. Kob, C. Donati, S. J. Plimpton, P. H. Poole, and S. C. Glotzer, Phys. Rev. Lett. {\bf 80}, 2827 (1997); C. Donati, J. F. Douglas, W. Kob, S. J. Plimpton, P. H. Poole, and S. C. Glotzer, Phys. Rev. Lett. {\bf 80}, 2338 (1998); C. Donati, S. C. Glotzer, P. H. Poole, W. Kob, and S. J. Plimpton, Phys. Rev. E {\bf 60}, 3107 (1999); S. C. Glotzer and C. Donati, J. Phys.: Condens. Matter {\bf 11}, A285 (1999).

\bibitem{mountain} D. Thirumalai and R. D. Mountain, J. Phys. C {\bf 20}, L399 (1987); R. D. Mountain and D. Thirumalai, Phys. Rev. A {\bf 36}, 3300 (1987); A. I. Mel'cuk, R. A. Ramos, H. Gould, W. Klein, and R. D. Mountain, Phys. Rev. Lett. {\bf 75}, 2522 (1995).

\bibitem{heuer} A. Heuer, Phys. Rev. Lett. {\bf 78}, 4051 (1997); S. Buchner and A. Heuer, Phys. Rev. E {\bf 60}, 6507 (1999); J. Qian, R. Hentschke, and A. Heuer, J. Chem. Phys. {\bf 110} 4514 (1999); {\it ibid} {\bf 111}, 10177 (1999). 

\bibitem{schober} C. Oligschleger and H. R. Schober, Phys. Rev. B {\bf 59}, 811 (1999); D. Caprion, J. Matsui, and H. R. Schober, Phys. Rev. Lett. {\bf 85}, 4293 (2000).

\bibitem{perera1} D. N. Perera and P. Harrowell, Phys. Rev. E {\bf 59}, 5721 (1999); J. Chem. Phys. {\bf 111}, 5441 (1999).

\bibitem{angelani1} L. Angelani, G. Parisi, G. Ruocco, and G. Viliani, Phys. Rev. Lett. {\bf 81}, 4648 (1998).

\bibitem{mukherjee} A. Mukherjee, S. Bhattacharyya, and B. Bagchi, J. Chem. Phys. {\bf 116}, 4577 (2002).

\bibitem{gotze} U. Bengtzelius, W. Gotze, and A. Sjolander, J. Phys. C {\bf 17}, 5915 (1984); W. Gotze and L. Sjogren, Rep. Prog. Phys. {\bf 55}, 241 (1992).

\bibitem{schroder} T. B. Schroder, S. Sastry, J. C. Dyre, and S. C. Glotzer, J. Chem. Phys. {\bf 112}, 9834 (2000).

\bibitem{angelani2} L. Angelani, R. Di Leonardo, G. Ruocco, A. Scala, and F. Sciortino, Phys. Rev. Lett. {\bf 85}, 5356 (2000).

\bibitem{hansen} H. Miyagawa, Y. Hiwatari, B. Bernu, and J. P. Hansen, J. Chem. Phys. {\bf 88}, 3879 (1988).

\bibitem{sarika1} S. Bhattacharyya, A. Mukherjee, and B. Bagchi, J. Chem. Phys. {\bf 117}, 2741 (2002).

\bibitem{sarika2} S. Bhattacharyya and B. Bagchi, Phys. Rev. Lett. {\bf 89}, 025504 (2002).

\bibitem{stillinger} J. A. Hodgdon and F. H. Stillinger, Phys. Rev. E {\bf 48}, 207 (1993).

\bibitem{kivelson} G. Tarjus and D. Kivelson, J. Chem. Phys. {\bf 103}, 3071 (1995).

\bibitem{liu} C. Z.-W. Liu and I. Oppenheim, Phys. Rev. E {\bf 53}, 799 (1996).

\bibitem{rossler} E. Rossler, Phys. Rev. Lett. {\bf 65}, 1595 (1990).

\bibitem{cicerone} M. T. Cicerone, F. R. Blackburn, and M. D. Ediger, J. Chem. Phys. {\bf 102}, 471 (1995); M. T. Cicerone and M. D. Ediger, {\it ibid.} {\bf 104}, 7210 (1996).

\bibitem{sillescu} F. Fujara, B. Geil, H. Sillescu, and G. Fleischer, Z. Phys. B {\bf 88}, 195 (1992); I. Chang {\it et al.}, J. Non-Cryst. Solids {\bf 172-174}, 248 (1994).

\bibitem{thirumalai} D. Thirumalai and R. D. Mountain, Phys. Rev. E {\bf 47}, 479 (1993).

\bibitem{yamamoto1} R. Yamamoto and A. Onuki, Phys. Rev. E {\bf 58}, 3515 (1998). 

\bibitem{yamamoto2} R. Yamamoto and A. Onuki, Phys. Rev. Lett. {\bf 81}, 4915 (1998).

\bibitem{perera2} D. N. Perera and P. Harrowell, Phys. Rev. Lett. {\bf 81}, 120 (1998).

\bibitem{leporini} C. D. Michele and D. Leporini, Phys. Rev. E {\bf 63}, 036701 (2001).

\bibitem{barrat} J. N. Roux, J. L. Barrat, and J. P. Hansen, J. Phys.: Condens. Matter {\bf 1}, 7171 (1989); 
J. L. Barrat, J. N.Roux and J. P. Hansen, Chem. Phys. {\bf 149}, 197 (1990).

\bibitem{wahnstrom} G. Wahnstrom, Phys. Rev. A {\bf 44}, 3752 (1991).

\bibitem{kob2} W. Kob and J. L. Barrat, Phys. Rev. Lett. {\bf 78}, 4581 (1997); K. Vollmayr-Lee, W. Kob, K. Binder, and A. Zippelius, J. Chem. Phys. {\bf 116}, 5158 (2002).

\bibitem{pusey} P. N. Pusey, {\it Liquids, Freezing and Glass Transitions, Les Houches session}, edited by J. P. Hansen, D. Levesque, and J. Zinn-Justin (North-Holland, Amsterdam, 1991); P. N. Pusey and W. van Megen, Phys. Rev. Lett. {\bf 59}, 2083 (1987); P. N. Segre, S. P. Meeker, P. N. Pusey, and W. C. K. Poon, Phys. Rev. Lett. {\bf 75}, 958 (1995).

\bibitem{russel} A. P. Gast and W. B. Russel, Phys. Today {\bf 51}, 24 (1998).

\bibitem{kofke} D. A. Kofke and P. G. Bolhuis, Phys. Rev. E {\bf 59}, 618 (1999).

\bibitem{lacks} D. J. Lacks and J. R. Wienhoff, J. Chem. Phys. {\bf 111}, 398 (1999).

\bibitem{sear} R. P. Sear, J. Chem. Phys. {\bf 113}, 4732 (2000).

\bibitem{doliwa} B. Doliwa and A. Heuer, Phys. Rev. Lett. {\bf 80}, 4915 (1998).

\bibitem{pham} K. N. Pham {\it et al.}, Science {\bf 296}, 104 (2002).

\bibitem{kiriushcheva} N. Kiriushcheva and P. H. Poole, Phys. Rev. E {\bf 65}, 011402 (2001).

\bibitem{allen} M. P. Allen and D. J. Tildesley, {\it Computer Simulation of Liquids}
(Oxford University Press, Oxford, 1987).

\bibitem{nose} G. J. Martyna, D. J. Tobias, and M. L. Klein, J. Chem. Phys. {\bf 101}, 4177 
(1994); H. C. Andersen, J. Chem. Phys. {\bf 72}, 2384 (1980); S. Nose, Mol. Phys. {\bf 52}, 255
(1984); W. G. Hoover, Phys. Rev. A {\bf 31}, 1695 (1985).

\bibitem{martyna1} G. J. Martyna, M. E. Tuckerman, and M. L. Klein, J. Chem. Phys. {\bf 97}, 2635
(1992).

\bibitem{tuckerman} M. E. Tuckerman, G. J. Martyna, and B. J. Berne, J. Chem. Phys. {\bf 97}, 1990
(1992).

\bibitem{martyna2} G. J. Martyna, M. E. Tuckerman, D. J. Tobias, and M. L. Klein, Mol. Phys.
{\bf 87}, 1117 (1996).

\bibitem{hm} J. P. Hansen and I. R. McDonald, {\it Theory of Simple Liquids} (Academic, London, 1986).

\bibitem{angel3} C. A. Angell, Science {\bf 267}, 1924 (1995).

\bibitem{murarka} R. K. Murarka, S. Bhattacharyya, and B. Bagchi, J. Chem. Phys. {\bf 117}, 10730 (2002).

\end{thebibliography}
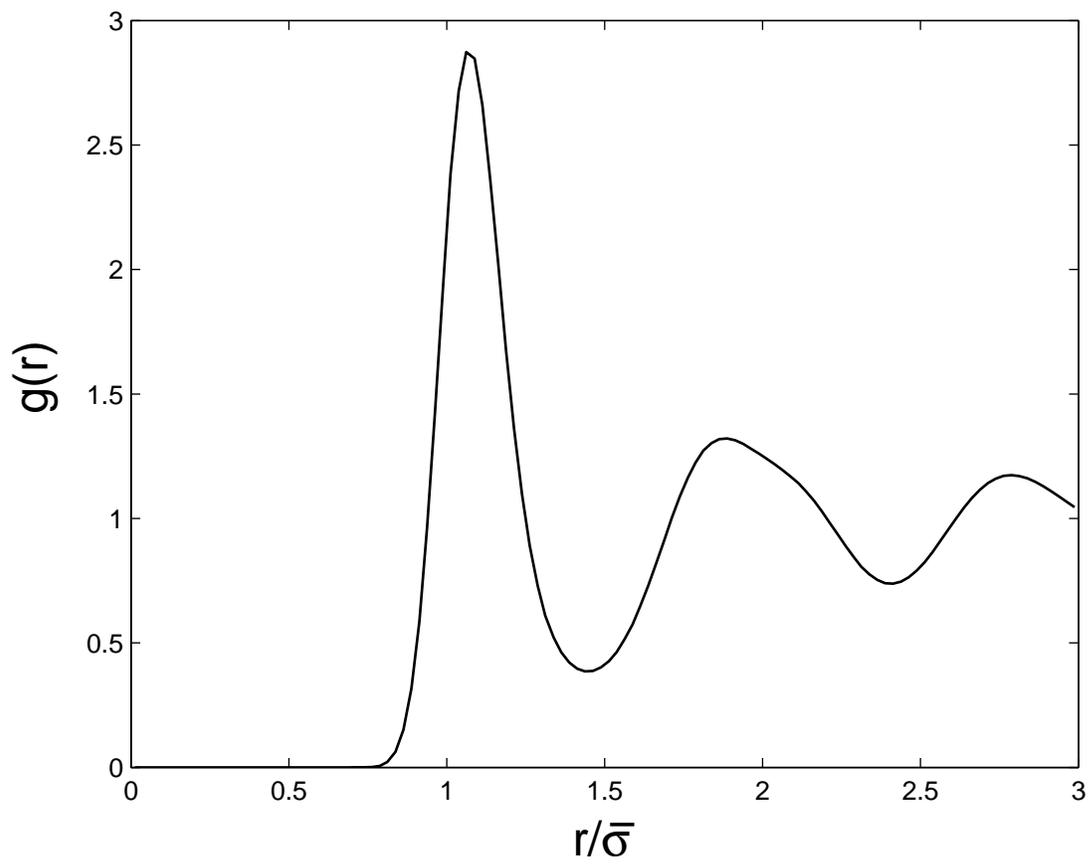
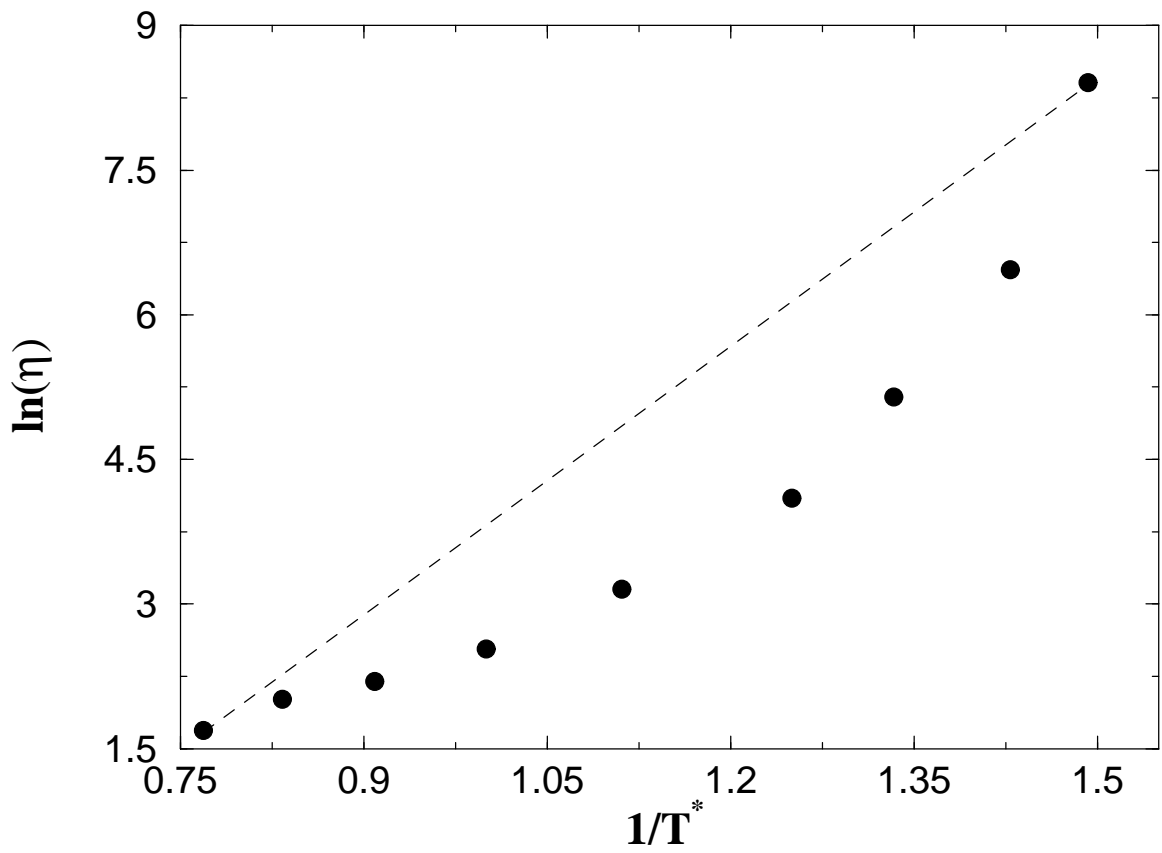
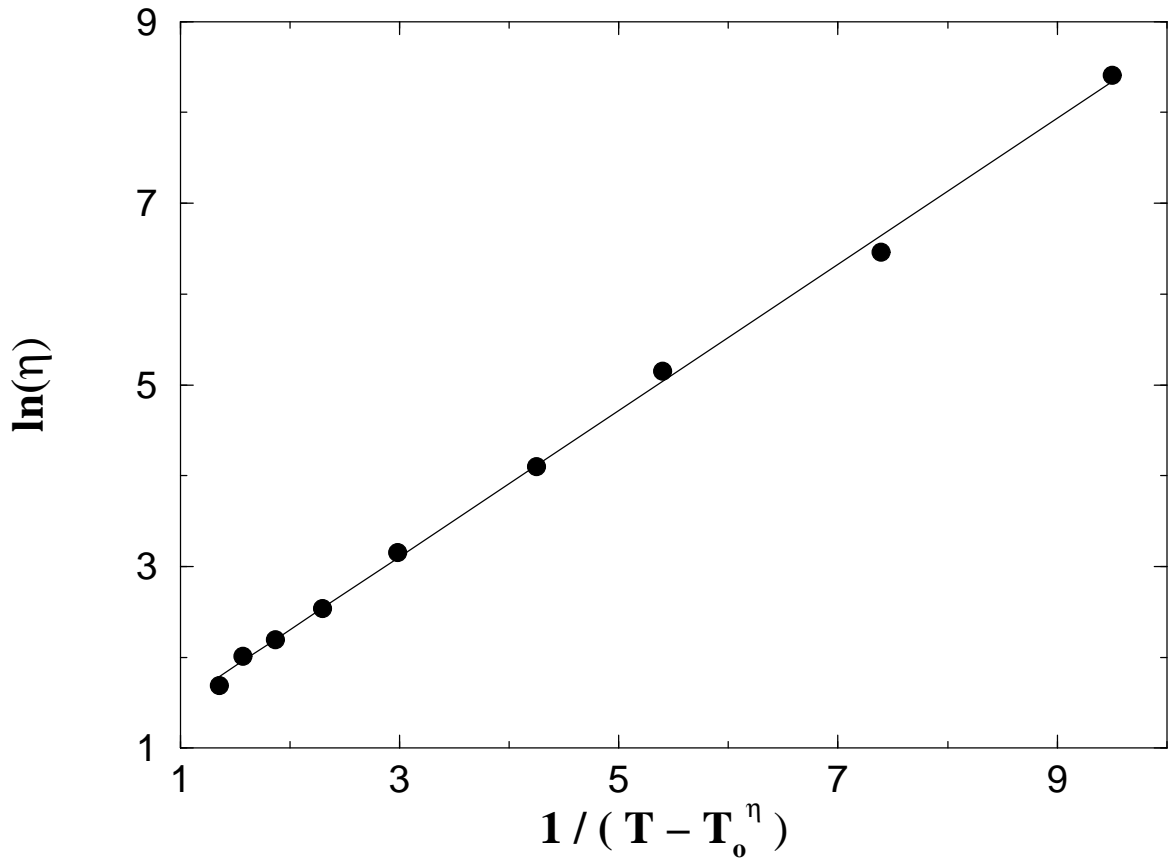
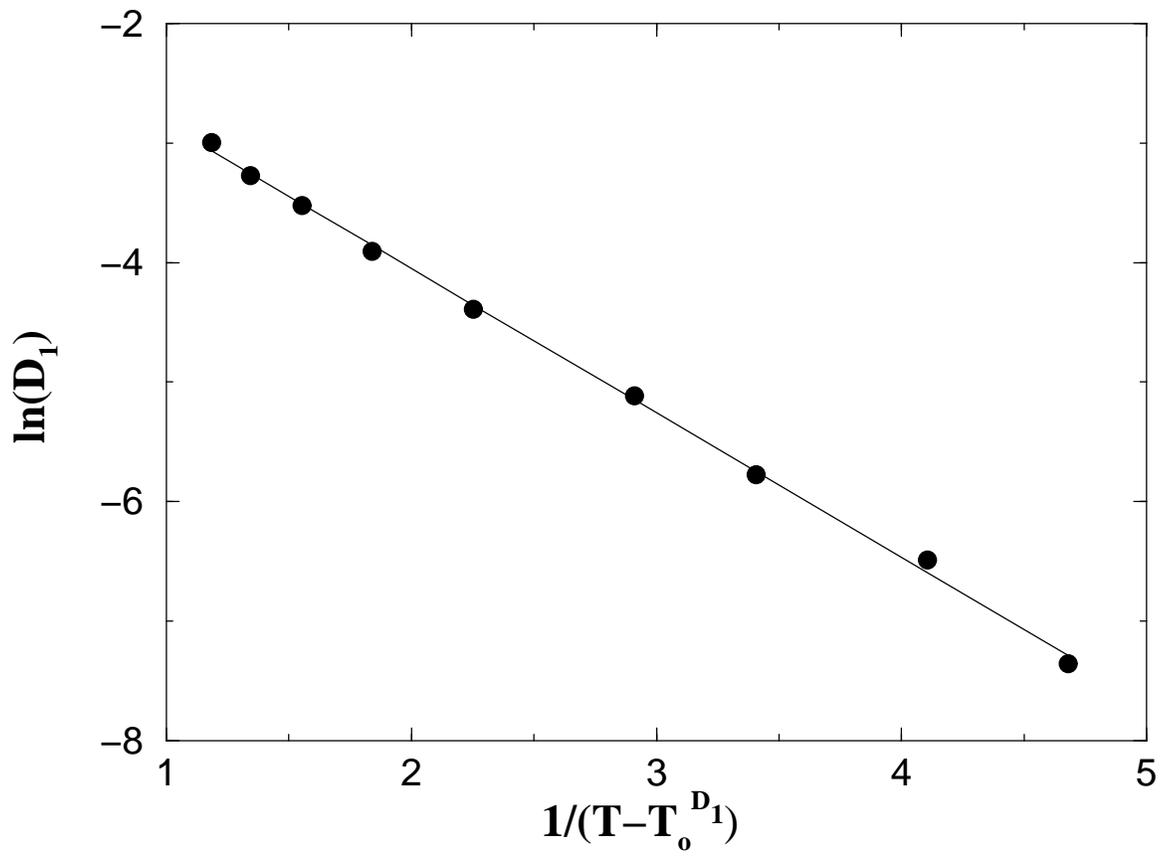
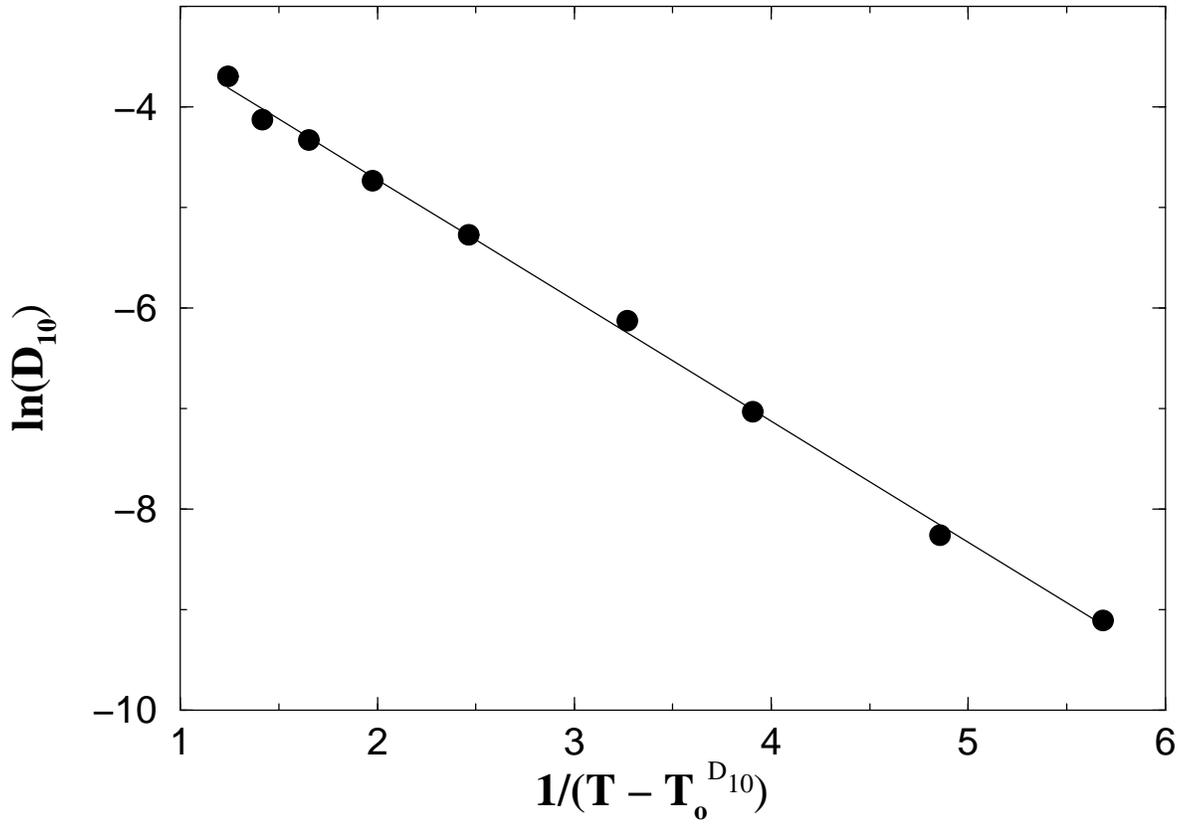
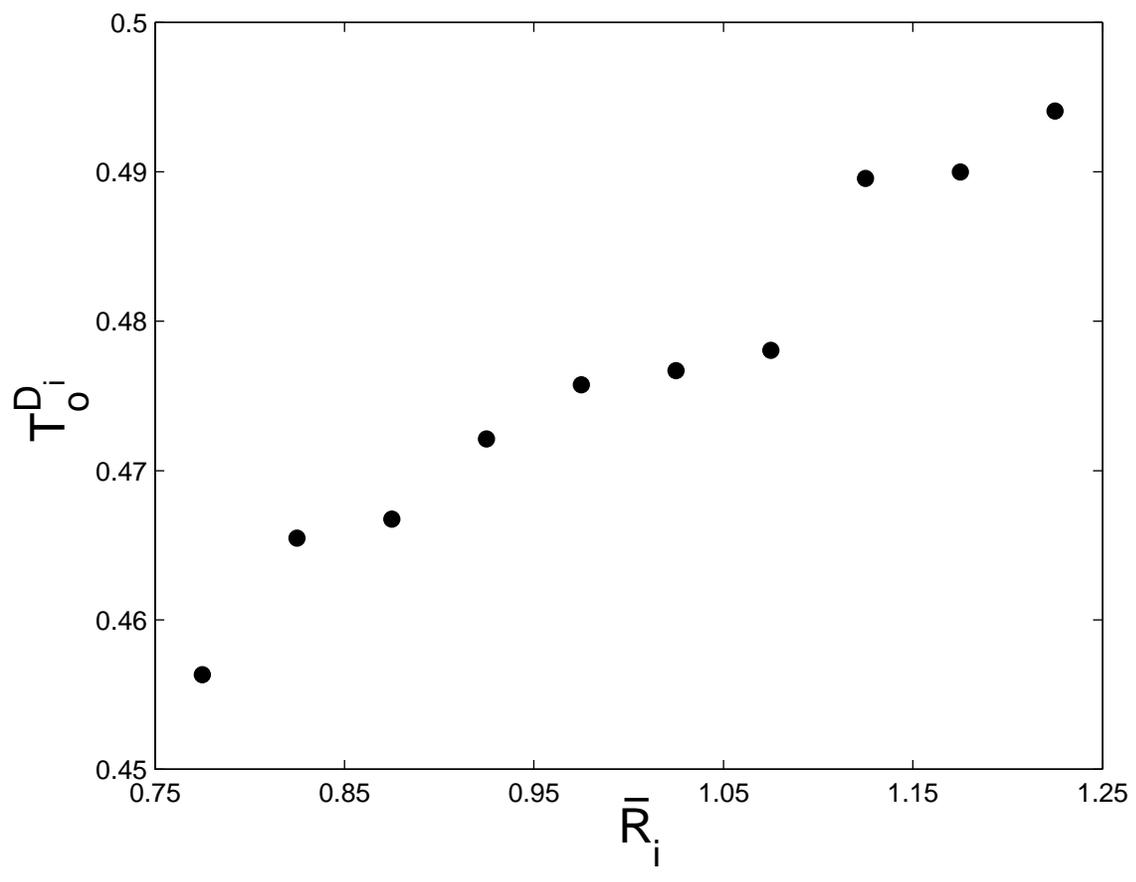
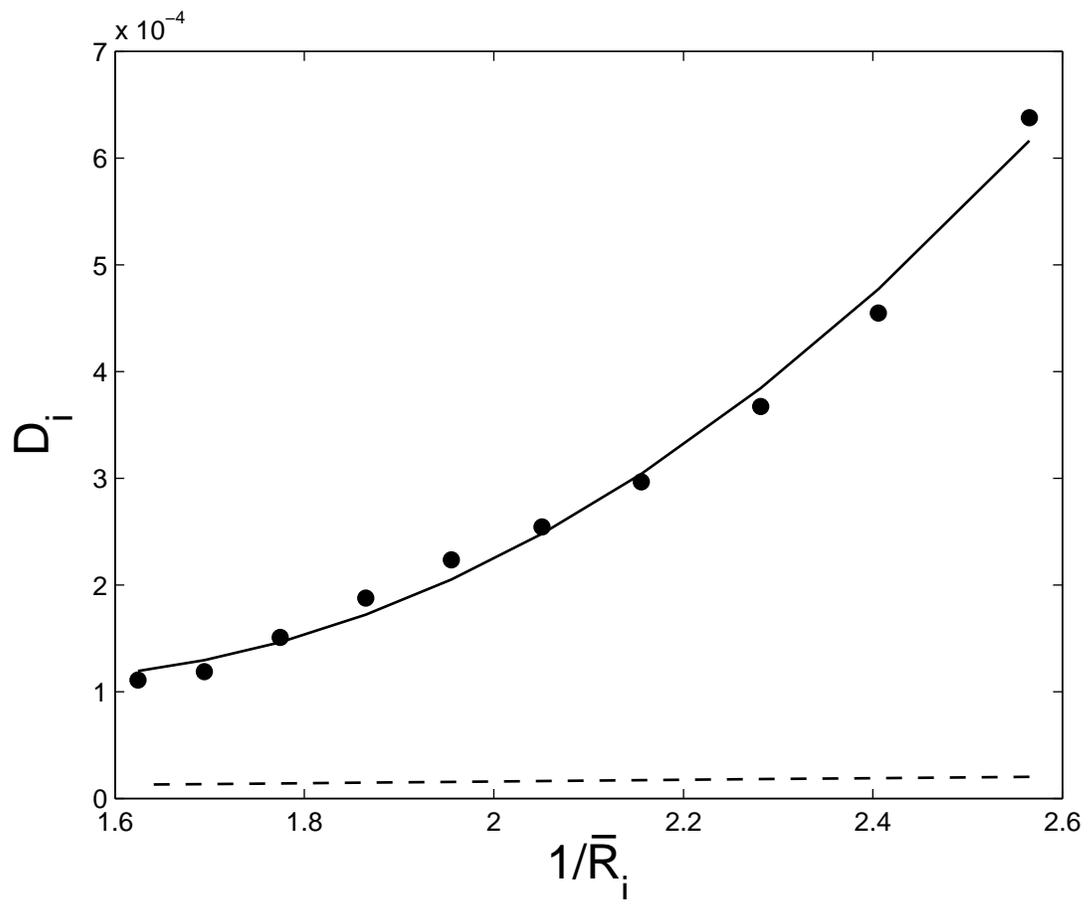
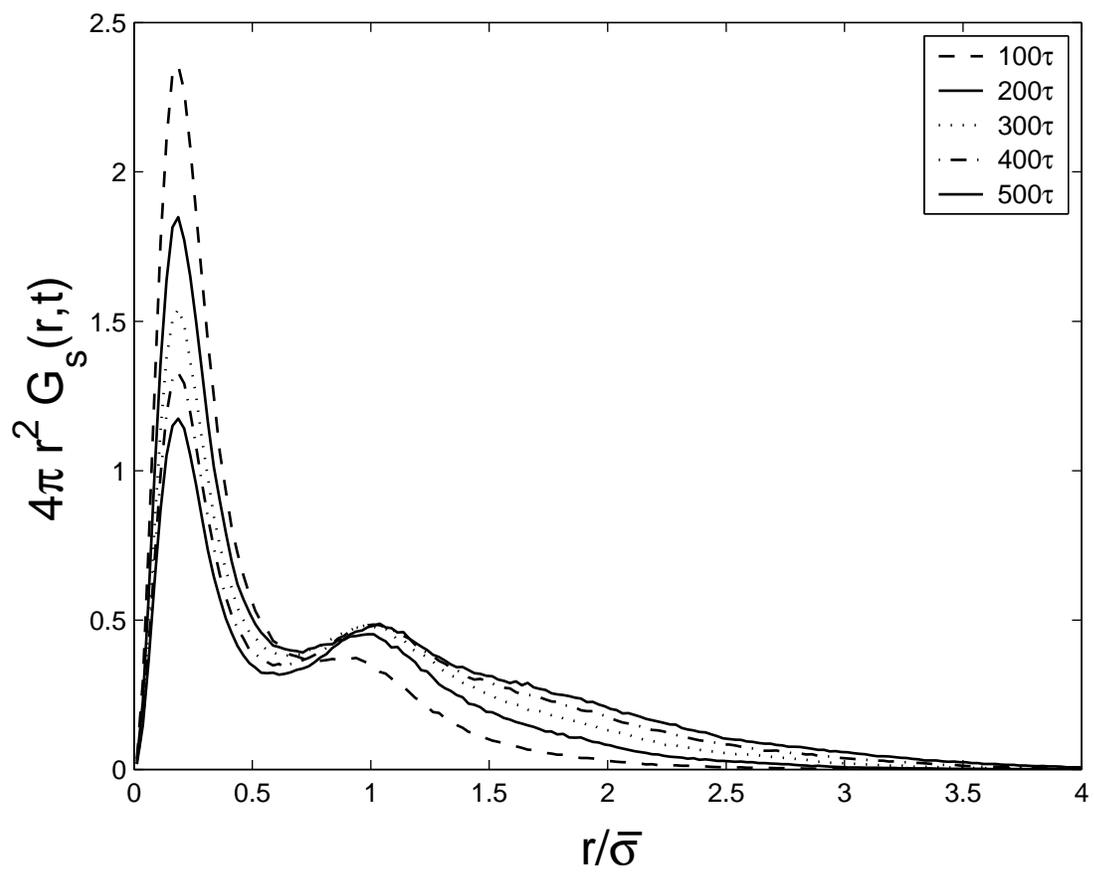
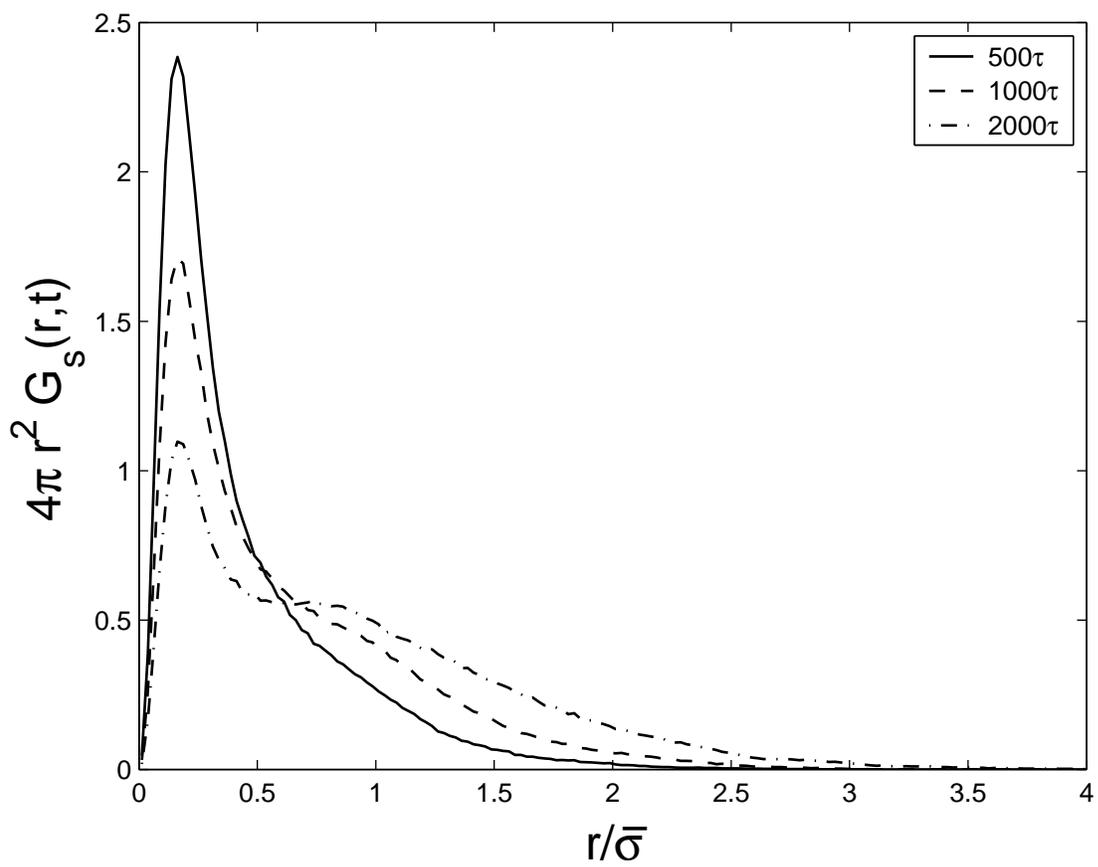
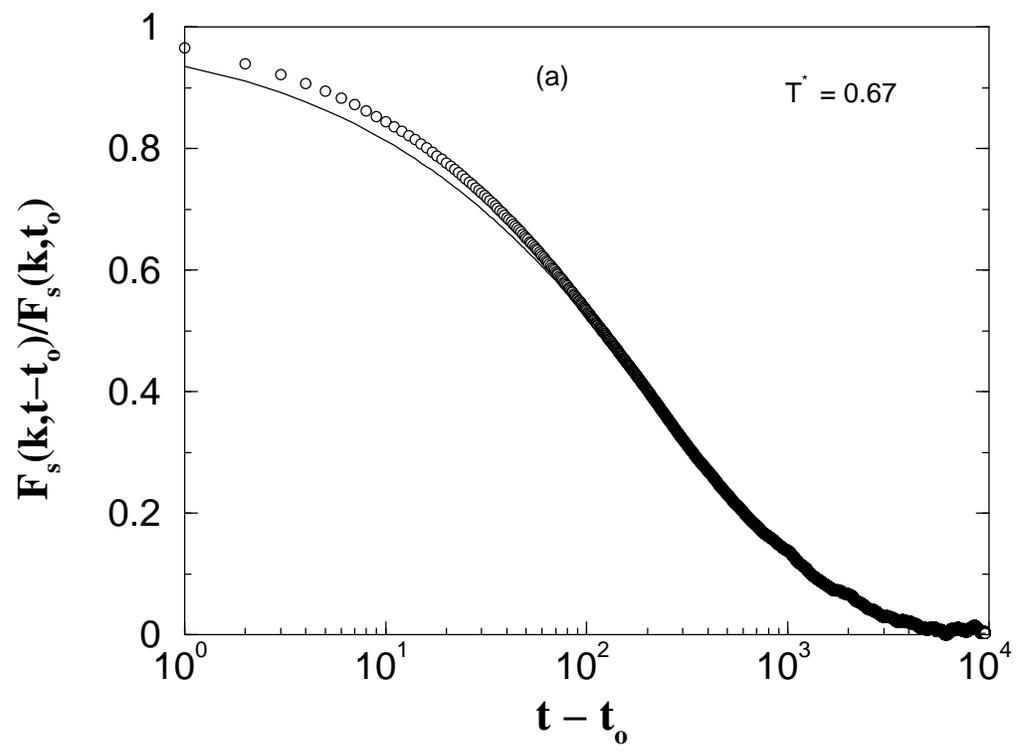
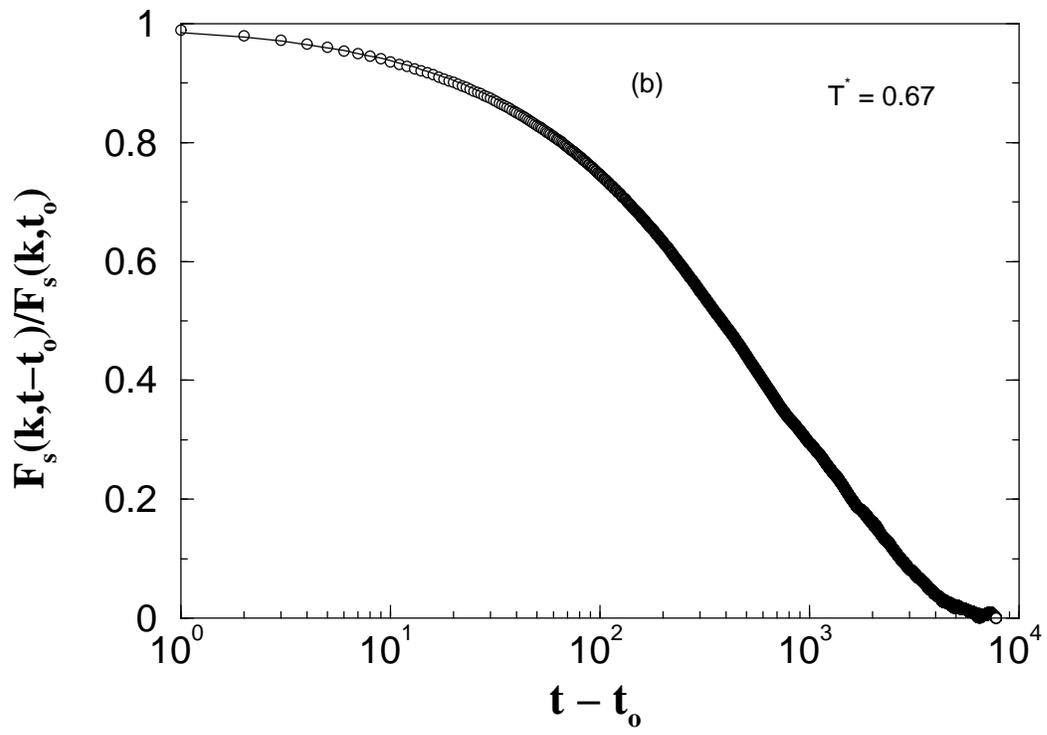
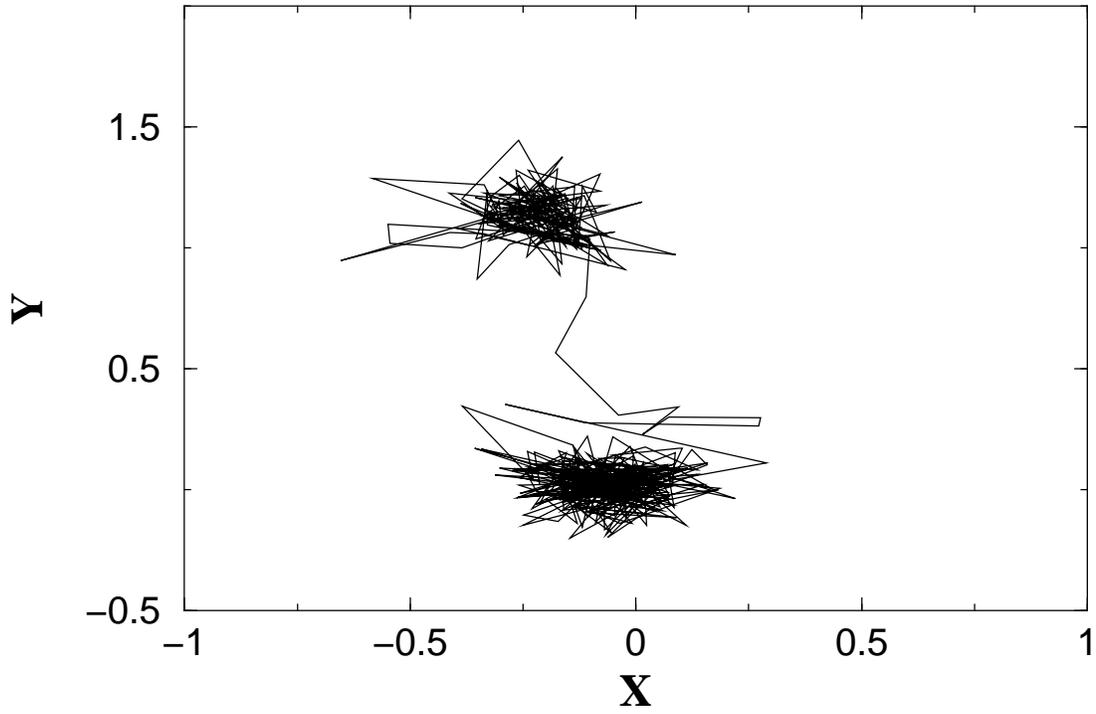
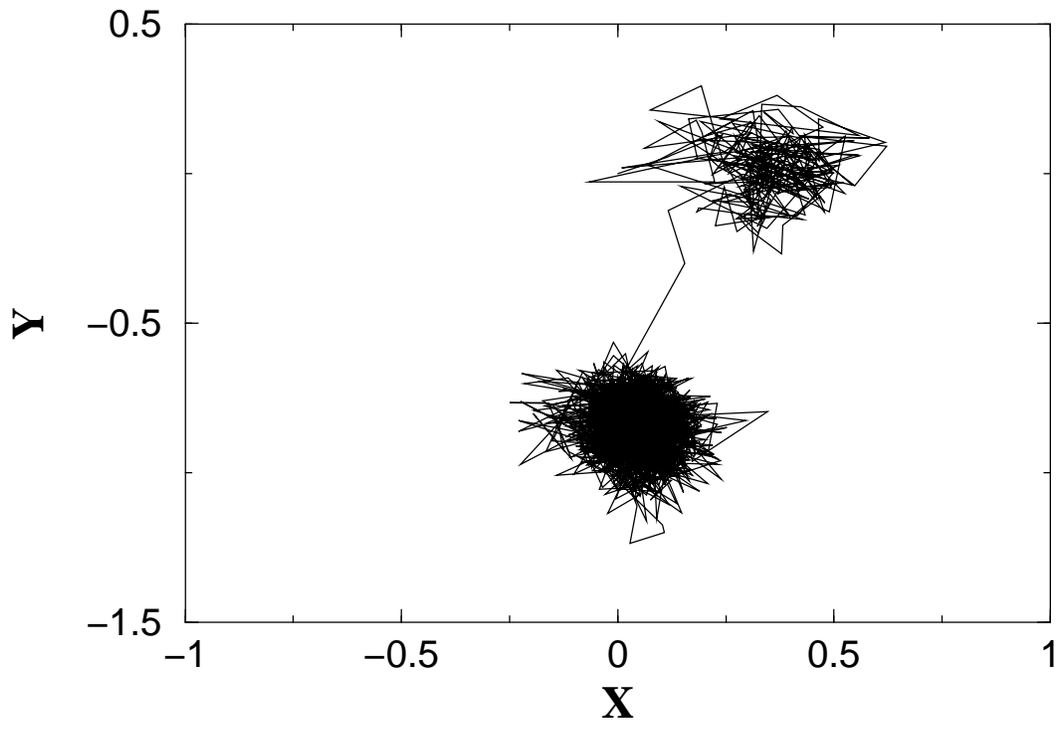
\end{document}